\documentclass[modern]{aastex63}
\usepackage{amsmath}
\usepackage{amssymb}
\usepackage{amsfonts}
\usepackage{amstext}
\usepackage{pstricks}
\usepackage{graphicx}
\usepackage{fancybox}
\usepackage{bm}
\usepackage{comment}
\usepackage{soul, color}
\usepackage{latexsym}
\usepackage{pifont}
\usepackage{shorttoc}
\usepackage{subfigure}
\usepackage{textcomp}
\usepackage{float}
\usepackage{enumitem}
\usepackage{longtable}
\usepackage{calrsfs}
\usepackage{bm}
\usepackage{lipsum}
\usepackage{threeparttablex}
\usepackage{hyperref}
\hypersetup{urlcolor=blue}
\def\aap{A\&A}

\def\apjs{ApJS}
\def\apjl{ApJL}

\DeclareMathAlphabet{\pazocal}{OMS}{zplm}{m}{n}

\newcommand{\no}[1]{}

\newcommand{\myemail}{lmr@ucsc.edu}
\def\lsim{~\rlap{$<$}{\lower 1.0ex\hbox{$\sim$}}}

\def\gsim{~\rlap{$>$}{\lower 1.0ex\hbox{$\sim$}}}

\newcommand{\rc}{\ensuremath{r_c}}
\newcommand{\dell}{\ensuremath{d_\ell}}
\newcommand{\zell}{\ensuremath{z_\ell}}
\newcommand{\dls}{\ensuremath{d_{\ell s}}}
\newcommand{\fv}{\ensuremath{f_v}}
\newcommand{\fvr}{\ensuremath{\fv(r)}}
\newcommand{\rtwo}{\ensuremath{r_{200}}}
\newcommand{\nimg}{\ensuremath{N_{\rm img}}}

\shorttitle{CGM cloud sizes from refractive FRB scattering}
\shortauthors{Mas-Ribas, McQuinn \& Prochaska}

\begin{document}

\title{\Large CGM cloud sizes from refractive FRB scattering}

\email{\myemail}

\author{Llu\'is Mas-Ribas} 
\affiliation{Department of Astronomy and Astrophysics, University of California,
1156 High Street, Santa Cruz, CA 95064, USA} 
\affiliation{University of California Observatories, 1156 High Street, Santa Cruz, CA 95064, USA}
\author{Matthew McQuinn}
\affiliation{Department of Astronomy, University of Washington, Seattle, WA 98195-1580, USA}
\author{J. Xavier Prochaska} 

\affiliation{Department of Astronomy and Astrophysics, University of California,
1156 High Street, Santa Cruz, CA 95064, USA} 
\affiliation{University of California Observatories, 1156 High Street, Santa Cruz, CA 95064, USA}
\affiliation{Kavli Institute for the Physics and Mathematics of the Universe (Kavli IPMU), 5-1-5 Kashiwanoha, Kashiwa, 277-8583, Japan}
\affiliation{Division of Science, National Astronomical Observatory of Japan, 2-21-1 Osawa, Mitaka, Tokyo 181-8588, Japan}

\begin{abstract}  

We explore constraints on the size of cool gas clouds in the circumgalactic medium (CGM) 
obtainable from the presence, or lack thereof, of refractive scattering in fast radio bursts 
(FRBs).  
Our refractive analysis sets the most conservative bounds on  
parsec-scale CGM clumpiness  as it 
does not make assumptions about the turbulent density cascade. We find that the bulk of low-redshift cool CGM gas, constrained to have densities of $n_{\rm e} \lsim 10^{-2}\,{\rm cm^{-3}}$,
likely cannot produce two refractive 
images and, hence, scattering. It is only for extremely small cloud sizes $\lesssim 0.1$ pc (about 
a hundred times smaller than the so-called shattering scale) that such densities could result in  detectable scattering.  Dense $n_{\mathrm e} \gtrsim 0.1\,{\rm cm^{-3}}$ gas with shattering-scale  
cloud sizes is more likely to inhabit the inner several kiloparsecs of the low-redshift CGM: such clouds would result in  multiple refractive 
images and large scattering times $\gtrsim 1 - 10$ ms, but a small fraction of FRB sightlines are likely to be affected. We argue that such large scattering 
times from an intervening CGM would be a signature of sub-parsec clouds, even if diffractive scattering from turbulence contributes to the overall scattering. At redshift $z\sim  3$, we estimate $\sim 0.1\%$ of FRBs to  
intersect massive proto-clusters, which may be the most likely place to see scattering owing to their ubiquitous $n_{\rm e} \approx 1\,{\rm cm^{-3}}$ cold gas. While much of our discussion assumes a single cloud size, we show similar results hold for a CGM cloud-size distribution motivated by hydrodynamic simulations.

\end{abstract}
  
\vspace{1.8cm}

\section{Introduction}\label{sec:intro}

Fast radio bursts (FRBs) are luminous millisecond-duration radio pulses of yet an unclear origin 
\citep{Petroff2022}, arising from a variety of galactic environments, and from dwarf to massive 
galaxies \citep[e.g.,][]{Mannings2021,Bhardwaj2023,Hewitt2024,Shah2024, Eftekari2024,Sharma2024}. 
Because FRBs are extragalactic sources, currently detected up to $z\gtrsim 1$ \citep{Ryder2023}, 
they can be used to probe the media intersected 
by their light on its way from the source 
to our telescopes  
\citep{Mcquinn2014,Prochaska2019,Macquart2020,Simha2020,Lee2022,Baptista2023,Simha2023,Lee2023,Connor2024,Ilya2024}. 

Of particular interest here is the potential to study the circumgalactic medium (CGM) in  intervening 
halos along the line-of-sight to FRBs. The CGM is the gaseous region extending out to a few hundreds 
of kiloparsecs around galaxies, connecting  
the intergalactic (IGM) and interstellar (ISM) media. It is 
a multiphase environment in which cool (${\rm T}\sim10^4$ 
K) gas clouds are embedded in a more diffuse and hotter 
(${\rm T}\sim10^6$ K) component, but the detailed features   
of this cool gas remain unknown \citep[see][for 
reviews]{Tumlinson2017,Faucherguigere2023}. Low-ionization metal 
transitions in the spectra 
of quasar sightlines intersecting the CGM indicate that the cool 
gas has an almost unity areal covering fraction \citep[][]{Chen2010,Nielsen2013,Werk2013,Dutta2020,Huang2021}, with 
a smaller ($\lesssim 10\%$) volume filling fraction 
\citep[][]{Stocke2013,Werk2014,Stern2016,Faerman2023}. Furthermore, 
the line width of these metal lines \citep[$\sim10 - 50\,{\rm 
km\,s^{-1}}$;][]{Werk2013,Qu2022} probes the velocity structure 
of the gas, but it is still unclear whether these widths are set by   
the turbulent velocity within a single cloud or from the velocity dispersion 
of many small independent clouds. 

The sizes of the CGM clouds 
carry information about the physical processes responsible 
for the multiphase structure, but this quantity is difficult to 
constrain observationally.  Sizes of absorption {\it systems} 
of $\sim 1 - 10$ kpc have been probed from multiply lensed quasars  
\citep[][and similarly 
from the sightlines to background galaxies; \citealt{Lopez2018,Peroux2018,Rubin2018b,Tejos2021}]{Rauch2001a,Ellison2004,Rubin2018c,Kulkarni2019,Augustin2021,Dutta2024}, 
and coherent scales down to a few hundred pc were inferred   
with quasar sightline 
pairs by, e.g., \cite{Rauch2001a} and \cite{Rudie2019}.
Furthermore, a median cloud size of $\sim 100$ pc in the CGM 
of (quiescent) luminous red galaxies was inferred by 
\citealt{Zahedy2019} (see also \citealt{Zahedy2021})  
through photoionization modeling, with their data 
covering the range $\sim 10 - 10^4$ pc. \cite{Chen2023} performed 
a similar analysis to that of \cite{Zahedy2019} including star-forming 
galaxies and  found  a wider range of sizes that extend to clouds 
of $\sim 1$ pc. Finally, it is worth noting the large 
population of so-called high-velocity clouds (HVC) inhabiting the inner  
$\sim 10$ kpc of the Milky Way halo \citep[][]{Wakker2008}, with 
a small fraction of them also found  at large Galactic 
distances (several tens of kpc), as well as in the CGM of Andromeda 
\citep{putman2003,Pisano2007}. HVCs are often  detected in 21 cm and H$\alpha$ 
emission, and they show self-similar (fractal) structure down to the 
spatial resolution of observations \citep{Vogelaar1994}, which may 
indicate very small cloud sizes \citep[see also][for reviews]{Wakker1997,Putman2011}.

From the theoretical side, there has recently been substantial work on the potential for parsec-scale  clouds at $\sim 10^4$ K condensing from hot gas. \citet{Mccourt2018} pointed out that a very natural scale for the gas to clump is the sound speed times the cooling time, at least in the plausible limit where conduction is highly suppressed by magnetic fields.  They found in hydrodynamic simulations that cooling gas tends to fragment on this `shattering' scale.  Others have found that even if the gas does not shatter, it may `splatter' and form cloudlets at the shattering scale if it reaches the $\sim 10^4$ K equilibrium temperature with supersonic inward velocities \citep{2023FrASS..1098135W, 2023MNRAS.525.1839F}.  While linear perturbations to virialized gas may not shatter because of thermal  instability \citep{2023FrASS..1098135W}, nonlinear perturbations in the virialized gas can break up on the shattering scale \citep{2020MNRAS.494L..27G}.  If formed, it is then possible that the small cloudlets re-coagulate into larger clouds as found in \citet{2021MNRAS.502.4935D} and \citet{Gronke2022}.  We will  consider the results of the simulations of \citet{Gronke2022} that find that such coagulation results in a power-law distribution of cloud sizes.

Various physical quantities that characterize the intervening CGM can be obtained from FRB 
observables. For example, the total column density 
of free electrons traversed by the radio signal, the so-called dispersion measure (DM), 
induces a delay in the arrival time of the photons that has a characteristic frequency 
dependence of $\nu^{-2}$. Thus, the modeling of this dependence allows us to infer the 
electron column of all the media intersected by the radio waves. When the dispersion measure is weighted by the parallel 
component of the magnetic field in the medium along the FRB sightline, one obtains 
the quantity referred to as Faraday rotation measure (RM), related to the polarization 
of the radiation. Furthermore, the scattering of photons 
by the intersected medium gives rise to additional effects: scattering 
 breaks up an FRB into multiple images, resulting in an 
observed angular broadening of the source. 
The arrival of these images at the observer location at different times (with 
time delays that depend on frequency as $\nu^{-4}$) results in 
their temporal superposition, which may yield to an overall extended  
signal dubbed scatter broadening \citep[][]{Williamson1972}. 
These multiple images can  also 
interfere with one another, an effect known as scintillation, where intensity  
fluctuations appear in the observed FRB spectrum. The existence and characteristics 
of the scatter broadening and scintillation signals depend on the physical properties 
of the intervening gas (its density and cloud sizes), as 
well as on its location along the sightline with respect to the FRB source and the 
observer \citep[e.g.,][see also \citealt{Ginzburg1970, Draine2011,Cordes2019}]{Narayan1992}. 
 Limits on the distance to the gas responsible for the scattering (addressing whether this gas is near the FRB source or further along the sightline) from the presence of scintillation from gas within the Milky Way  
have been addressed in previous work \citep[e.g.,][]{Masui2015,Sammons2023,Nimmo2024}, while here we will 
 focus on the constraints on the cloud sizes and density 
of the intervening CGM gas.

Scattering of radio waves is often categorized by taking into account the size of the 
scatterers.  Diffractive scattering comprises the regime where the scatterers are below 
the Fresnel scale and thus lead to light taking multiple interfering paths.  This is generally thought to occur from density fluctuations 
driven by turbulence \citep[e.g.,][]{Rickett1990}, as turbulence can lead to a cascade that reaches extremely small length scales. Turbulence is the canonical source for scattering in 
the Milky Way interstellar medium \citep{Armstrong1995}, although it is likely that other interstellar structures lead to at least some of the more extreme scattering events \citep{10.1093/mnras/stu1020}. For the 
case of a CGM at a distance of one gigaparsec, the scale of fluctuations below which scattering becomes diffractive is approximately an astronomical unit \citep{Prochaska2019}. Turbulence could lead to fluctuations on such small scales and indeed is observed to have structure on thousands of kilometer scales in the ISM. 
However, the details of the turbulence that can lead to fluctuations on diffractive scales is hotly debated \citep[e.g.][]{2022JPlPh..88e1501S}. Even in the context of \citet{1995ApJ...438..763G} MHD turbulence, \citet[see the supplementary documents]{Prochaska2019} stresses that various physical 
processes can result in the cut-off of the density spectrum in the density cascade prior to the 
diffractive scale.  For example, turbulent density fluctuations may not be passively advected by the flow to small eddies if the cascade becomes isothermal prior to eddies where transport is set by magnetic waves, and there may additionally be a cutoff from proton free streaming/diffusion that can be larger than the diffractive scale  \citep{Lithwick_2001}. Such cut-offs in the turbulent cascade 
would suppress or eliminate diffractive scattering.    

Refractive scattering, on the other hand, considers scatterers larger than the Fresnel scale. It is particularly sensitive to discrete CGM clouds of sizes of $\,\lesssim 1$~pc \citep{Vedantham2019}. As discussed above, these sizes   
are  challenging to probe with other observational methods, but 
they are potentially consistent with absorption-line spectroscopy results if each absorption system is composed of thousands of parsec-scale clouds 
\citep[e.g.,][]{Werk2016,Mcquinn2018,Faerman2024} as has been conjectured \citep{Mccourt2018}.  Refraction is also interesting because it potentially leads to the largest scattering times: We will argue 
that it is challenging to produce diffractive scattering from CGM turbulence that is significantly larger than our estimates for refractive scattering from clouds.

Finally, we point out that for the low-$z$ CGM, the typical 
inferred densities are such that the shattering scale is likely tens or 
hundreds of parsecs.  Thus, even if the CGM does shatter and result in such cloud sizes, these clouds are too large to produce refractive scattering. This contrasts with the findings of previous studies, which suggest that FRB scattering places strong limits on low-redshift shattering-scale cloudlets \citep{Vedantham2019, Jow2023}.

In this paper, we investigate the constraints on the size of cool CGM gas 
clouds that can be obtained from refractive scattering, and discuss these 
effects and results in the context of actual CGM observations. 
Section~\ref{sec:theory} presents the refraction formalism, adopting 
part of the methods in \cite{Jow2023} who explored similar constraints 
for sub-parsec cloudlets via a combined plasma and lensing approach. 
A fiducial toy model halo 
is used in Section~\ref{sec:ensemble} to infer dependencies on halo parameters and to obtain general results. 
In Section~\ref{sec:ensemblediff}, we explore variations of the fiducial halo and alternate 
physical scenarios motivated by observations and simulations. We compare the 
contributions of refractive and diffractive CGM scattering in 
Section~\ref{sec:turb}, and conclude in Section~\ref{sec:conclusions}.

\section{Formalism}\label{sec:theory}

Section~\ref{sec:formalism} below describes the formalism for refractive scattering  in the circumgalactic medium of 
extragalactic halos. The impact of extragalactic refractive scattering on Milky Way scintillation  is then presented 
in Section~\ref{sec:scintillation}.

\subsection{Refractive Scattering}\label{sec:formalism}

The (rest-frame) plasma frequency is 
\begin{equation}\label{eq:wp}
    \nu_p = \sqrt{\frac{e^2 n_{\rm e}}{4 \pi^2 \epsilon_0 m_{\rm e}}},
\end{equation}
where $m_{\rm e}$ and $e$ are the electron mass and charge, respectively, $n_{\rm e}$ is 
the electron density of the medium, and $\epsilon_0$ is the vacuum permittivity\footnote{The term 
$4 \pi \epsilon_0$ in the denominator equals unity in cgs units often used in astronomy. We explicitly write this term (in SI units) 
in all our expressions for clarity.}. Regions 
in the plasma with fluctuations $\Delta n_{\rm e}$ departing from the mean electron density translate 
into root mean square fluctuations of the refractive index \citep[][chapter 11]{Draine2011}
\begin{equation}\label{eq:deltam}
    \Delta m_{\rm rest} =  \frac{1}{2} \frac{\omega_{\rm p}^2}{\omega_{\rm rest}^2} = \frac{e^2 }{2 \epsilon_0 m_{\rm e} 
    \omega_{\rm rest}^2} \Delta n_{\rm e} ~  ,
\end{equation}
which induce phase velocity variations $\Delta v_{\rm phase} = c \Delta m_{\rm rest}$ of the radio waves 
propagating through such a medium. Here and in the
following, 
$\omega_{\rm rest} = 2 \pi \nu_{\rm rest}$ ($\omega_{\rm p} = 2 \pi \nu_{\rm p}$) is the rest-frame 
angular frequency of the radio waves (plasma) and $c$ denotes the speed of light.  
These effects, in turn, result in a shift of the wave phase 
\begin{equation}
    \Delta x_{\rm c} =  l \Delta m_{\rm rest} ~ ,
\end{equation}
where $l = 2\,r_{\rm c}$ is the path length that the radio waves traverse when encountering one 
region (cloud), which we will consider spherical with
radius \rc. 
Here, $l$ implies that the region is intersected 
through its center.

For pairs of rays 
separated by a distance $2\, r_{\rm c}$, such that they have independent paths, the difference in 
their phases results in a distortion of the wavefront, corresponding to a change in the direction 
of propagation of the radio waves by an angle \citep[see Figure 11.6 in][for 
a schematic representation]{Draine2011}
\begin{equation}
    \alpha_{\rm c} \approx g_{\rm s} \frac{\sqrt{2}\,\Delta x_c}{2\, r_{\rm c}} =  g_{\rm s}\sqrt{2}\,\Delta m_{\rm rest}   ~,
\end{equation}
where $g_{\rm s}=2.2$ is a geometric  
correction factor computed by \cite{Prochaska2019} that applies for 
perfectly spherical clouds\footnote{For slab-like clouds 
with the same characteristic $r_{\rm c}$ size, the correction term would simply 
equate unity, thus spherical clouds represent an upper limit for the angle (and 
scattering time) values.}.

In a medium consisting of many clouds, such as a galaxy halo, this angle is  
\begin{equation}\label{eq:deltax}
    \alpha \approx    \sqrt{\sum_j^{N_{\rm c}}\alpha^2_{{\rm c,}j} } =  g_{\rm s} \sqrt{2 N_{\rm c}}\, \Delta m_{\rm rest}   ~   ,
\end{equation}
where $N_{\rm c}$ is the number of clouds 
intersected by a ray and, since we now explicitly account for the number of 
clouds in the expressions, the term $\Delta m_{\rm rest}$  simply refers to a single cloud of density $n_{\rm e}$. 

With the expressions derived above and considering cosmological distances, we can now define 
the scattering angle in the observer frame as \citep[e.g.,][]{Macquart2013}
\begin{equation}\label{eq:theta}
    \Theta_s \approx  \frac{\dls}{d_{\rm s}} \alpha  = \frac{1}{(1+\zell)^{2}} 
    \frac{\dls}{d_{\rm s}} g_{\rm s} \sqrt{2 N_{\rm c}}\, \Delta m  ~ ,
\end{equation}
and the (geometric) scattering time, i.e., the time delay between images formed by refraction,  as 
\begin{equation}\label{eq:tau}
    \tau_s \approx (1+\zell) \frac{\dell d_{\rm s}}{\dls}\frac{\Theta_s^2}{2c}  =  \frac{1}{(1+\zell)^3}\frac{\dell \dls}{ d_{\rm s}} g_{\rm s}^2 {\frac{N_{\rm c}}{  c}} \Delta m^2 ~ .
\end{equation}
Here, \dell, $d_{\rm s}$, $d_{\rm \ell s}$ are the angular diameter distances \citep{Peacock1999} from the observer 
to the scattering screen (owing to plasma lensing from an intervening CGM), from the observer to the source, and from the scattering screen to the source, 
respectively. The term \zell\ denotes the redshift of the screen and $\Delta m$ is 
now defined in the observer frame via the observed frequency $\nu (1+\zell) = \nu_{\rm rest}$. Equation~\ref{eq:tau}  
highlights the fundamental role played by the number of clouds in the time delay. Specifically,   
it is the crossing of the radio waves into and out of these higher density media that induces 
scattering (refraction) and, therefore, the larger the number of clouds, the higher the scattering.

As we will show below, another relevant quantity in our work is 
the number of images produced by refractive scattering in the CGM (cf. \citealt{Prochaska2019, Jow2023}).  
We can define this term following \cite{Jow2023} as 
\begin{equation}\label{eq:nimg}
    N_{\rm img} \sim 1 + N_{\rm c} k_{\rm c}^2 ~,
\end{equation}
where  $k_{\rm c}$ is the lens convergence for a single cloud:
 \begin{equation}\label{eq:conv}
    k_{\rm c} =  \frac{1}{(1+z_{\rm l})^2}\frac{d_{\rm l}d_{\rm ls}}{ d_{\rm s}} {\frac{\sqrt{2}\,\Delta m}{ r_{\rm c}}} \sim \frac{\Theta_{\rm c}}{\theta_{\rm c}}  ~, 
\end{equation} 
where $\theta_{\rm c}=r_{\rm c}/\dell$ and $\Theta_{\rm c}$ are 
the angular scale and the scattering angle from a single CGM cloud 
in the intervening halo, respectively. 
Equation~\ref{eq:nimg} shows that even intersecting a single cloud 
($N_{\rm c}=1$) can result in more images than the original image of the source, as 
long as the cloud 
convergence is $k_{\rm c}\ge 1$.  When the number of clouds is large, as is the case of a halo, however, this condition is relaxed as then only the total column of clouds ($N_{\rm c} k_{\rm c}^2$) must be greater than unity to produce multiple images.

\subsection{Extragalactic Halos and their Effect on Milky Way Scintillation}\label{sec:scintillation}

One way to probe the impact of refractive scattering from an intervening 
extragalactic halo is on
the FRB scintillation signal created by the Milky Way. We briefly summarize here the aspects 
of this process that are relevant for our work, and refer the reader to 
\citealt{Jow2023} and \citealt{Pradeep2025}, for additional discussion and formulation (see also these works for schematic representations of 
these processes). 

Let us first consider scintillation as the result of multipath propagation arising from scattering 
in the Milky Way \citep[][]{Rickett1970}. In this scenario, multiple images created by scattering in 
the Milky Way, 
and separated by a characteristic time $\tau_{\rm MW}$ smaller than the pulse duration, interfere at 
the observer location giving rise to intensity fluctuations in the FRB spectrum with a characteristic 
frequency scale of $1/\tau_{\rm MW}$.

The presence of a distant circumgalactic medium (whether intervening or around 
the FRB host itself) along the FRB sightline that scatters the signal and 
creates multiple (i.e., more than one) images with a time delay $\tau_{\rm s}$, 
however, may eliminate the Milky Way scintillation signal. 
For this to happen, the Milky Way scattering 
screen must resolve the images from the extragalactic CGM \citep[i.e., 
each image from the halo must appear separated by an angle greater than 
$\sim \lambda/\sqrt{c\tau_{\rm MW} d_{\rm MW}}$, where $d_{\rm MW}$ is the distance to the Milky Way scattering screen and $\lambda = c\, \nu^{-1}$;][]{Sammons2023}, otherwise 
there would be no interference. By using that  the  
separation between images at the CGM lens is $\sqrt{\tau_{\rm s} c/d_{\rm \ell}}$, one can then obtain the lower limit for  $\tau_{\rm s}$ that 
enables the MW to resolve the CGM images given $\tau_{\rm MW}$:\footnote{
Here we have omitted cosmological distances for simplicity. Formal 
expressions accounting for this effect are derived in \cite{Pradeep2025} and appendix B in \cite{Sammons2023}.
}
\begin{equation}
    \tau_{\rm s} = 1\,\mu{\rm s}\left(\frac{d_{\rm \ell}}{1\, {\rm Gpc}}\right)\left(\frac{d_{\rm MW}}{1\, {\rm kpc}}\right)^{-1}
    \left(\frac{\tau_{\rm MW}}{1\, {\rm \mu s}}\right)^{-1}\left(\frac{\nu}{1\, {\rm GHz}}\right)^{-2}~. 
\end{equation} 
The larger the scattering 
time in the halo compared to the value required from the above expression, the more suppression of the Milky Way scintillation signal \citep{Pradeep2025}. We stress that the number of images must be 
$N_{\rm img} \ge2$ to have a defined scattering time and for this effect 
to happen.

In the following sections, we will use the expressions for the time delay and number of 
images  derived above to explore the constraints that can be obtained for intervening 
halos and the contribution of refraction to the overall scattering signal.

\section{A Toy Model Halo of Single-size Clouds}\label{sec:ensemble} 

We consider here the case of an extragalactic halo consisting of an 
ensemble of cool clouds of the same size and density, embedded in a more diffuse and 
warmer medium. These clouds represent the cool CGM medium 
that is observed in absorption in quasar spectra, and we 
ignore the contribution of the warm component assuming that 
its density, and thus scattering, is much 
lower than that of the cool gas \citep{Ocker2025}. This simple approach will allow us to gain 
broad insight into the dependence of the refractive scattering time delay 
and number of images on the cloud size and other halo parameters, 
as well as into the impact of scattering on  
scintillation and/or pulse broadening. 

We begin by considering a CGM with 
a constant radial volume filling fraction of cool clouds extending out 
to the halo virial radius $r_{200}$, a model that describes observations of the low-$z$ CGM \citep[e.g.,][]{Faerman2023}\footnote{These authors assume a constant filling factor but the density of their clouds 
depends on radial distance as $r^{-1.67}$, while we assume 
a constant density for simplicity.}: 
  \begin{equation}
    \fvr=
    \begin{cases}
      \fv & \text{for}\ r\leq\rtwo;  \\
      0 & \text{otherwise ~.}
    \end{cases}
  \end{equation}
The number of clouds intersected by a sightline traversing this halo 
at impact parameter $b$ is then 
\begin{equation}
N_{\rm c} (b)= 2\int_{0}^{\sqrt{r_{200}^2 -b^2}} \frac{\pi r_{\rm c}^2}{\frac{4}{3}\pi r_{\rm c}^3} f_{\rm v}\left(\sqrt{s^2+b^2}\right) \,{\rm d}{ s} = \frac{3}{2}  \frac{f_{\rm v}}{r_{\rm c}} \sqrt{r_{200}^2 -b^2} ~, 
\end{equation}
and if we further specialize to $b \ll \rtwo $:
\begin{equation}
N_{\rm c}(b)   \equiv N_{\rm c}= \frac{3}{2} f_{\rm v} \frac{\rtwo}{r_{\rm c}} ~. 
\end{equation}

With this expression, the dependence of the refractive scattering time delay on halo parameters 
(in addition to the halo redshift and its location relative to the observer 
and the source, as well as observed frequency) becomes 
\begin{equation}\label{eq:prop}
    \tau_s \propto  \frac{1}{(1+\zell)^3}\frac{\dell \dls}{ d_{\rm s}} f_{\rm v} \frac{ r_{200}} {r_{\rm c} } \frac{n_{\rm e}^2}{\omega^4} ~ .
\end{equation}
This expression indicates that the time delay from refractive scattering is proportional to the
size of the halo, the volume 
filling fraction, 
the square of the electron 
density in the clouds,  
and inversely proportional to the 
cloud size.
For the number of (additional) images created by refraction, there is 
a stronger dependence on the geometrical term and, most important, on the cloud size, compared to 
the time delay case, such that 
\begin{equation}\label{eq:propn}
    N_{\rm img} - 1 \propto \frac{1}{(1+\zell)^4} \left(\frac{\dell \dls}{d_{\rm s}}\right)^2 
    f_{\rm v} \frac{r_{200}}{r_{\rm c}^3} \frac{ n_{\rm e}^2}{w^4} ~.
\end{equation}

Adopting values for a fiducial halo and its CGM (motivated
by absorption-line studies), 
one estimates that the typical refractive time delay between images is
\begin{align}\label{eq:tauimage}
    \tau_{\rm s}  \approx \frac{2.5}{(1 + z_{\rm l})^{3}} {\rm ms} & 
    \left(\frac{\dell \dls/d_{\rm s}}{1\,{\rm Gpc}}\right)
    \left(\frac{{\rm \fv}}{0.1}\right)
    \left(\frac{{\rm r_{\rm 200}}}{200\,{\rm kpc}}\right)\nonumber \\
    & \times   \left(\frac{ n_{\rm e}}{10^{-2} \, {\rm cm^{-3}}}\right)^2 
    \left(\frac{r_{\rm c}}{1 \, {\rm pc}}\right)^{-1}
    \left(\frac{\nu}{1 \, {\rm GHz}}\right)^{-4} ~ \hspace{1cm} \textrm{ if \hspace{1cm} $N_{\rm img}\geq2$} ~, \nonumber \\
    & \hspace{7.5cm}  
\end{align}
and the number of images: 
\begin{align}\label{eq:imagetau}
    N_{\rm img} - 1 \approx \frac{10^{-2}}{(1 + z_{\rm l})^{4}} &
    \left(\frac{d_{\rm l}d_{\rm ls}/d_{\rm s}}{1\,{\rm Gpc}}\right)^2 \nonumber \\
    &\times \left(\frac{{\rm f_{\rm v}}}{0.1}\right)
    \left(\frac{{\rm r_{\rm 200}}}{200\,{\rm kpc}}\right)
    \left(\frac{ n_{\rm e}}{10^{-2} \, {\rm cm^{-3}}}\right)^2 
    \left(\frac{r_{\rm c}}{1 \, {\rm pc}}\right)^{-3} 
    \left(\frac{\nu}{1 \, {\rm GHz}}\right)^{-4} ~.
\end{align}

\begin{figure}\center 
\includegraphics[width=0.75\textwidth]{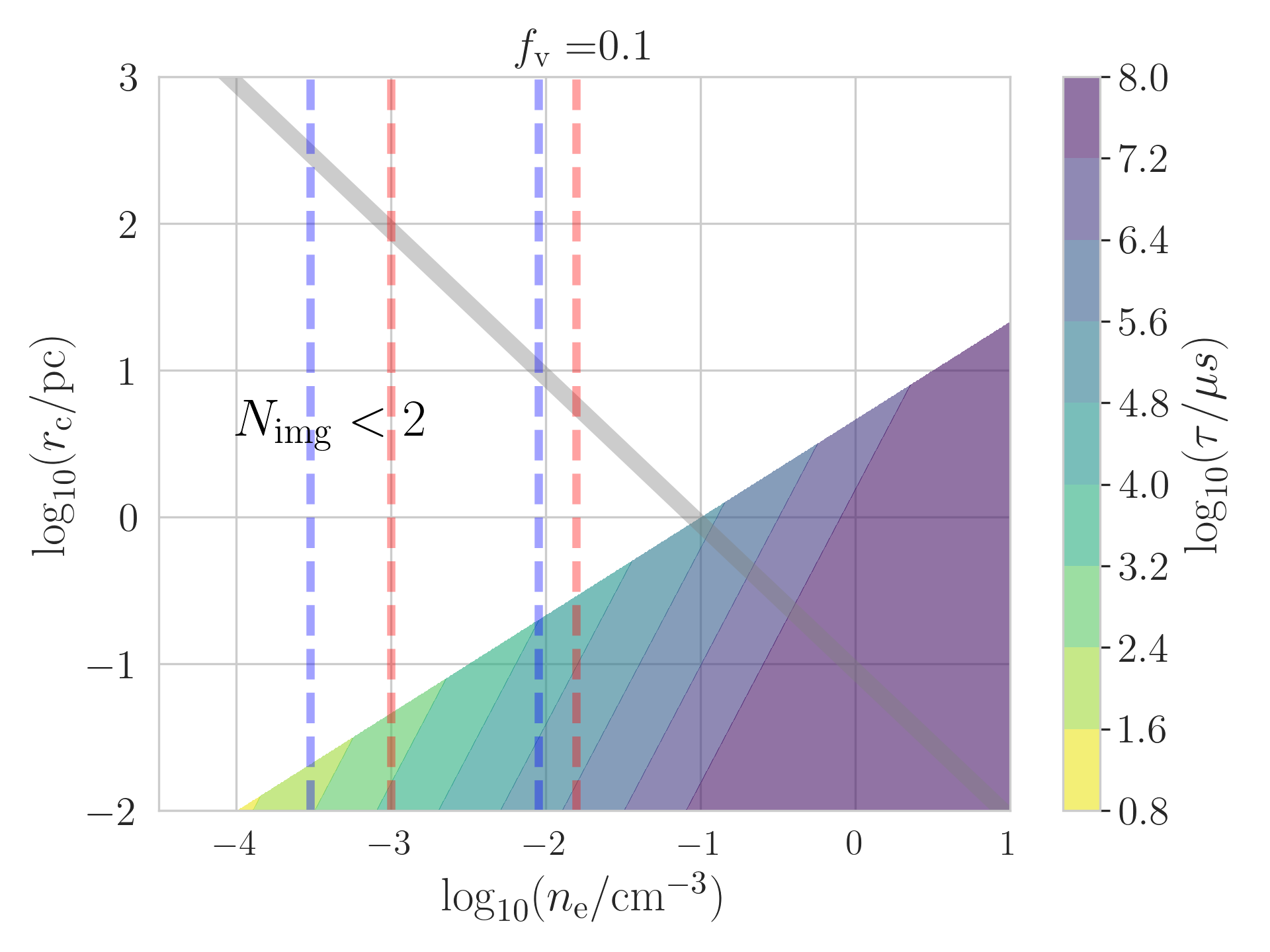}
\includegraphics[width=0.75\textwidth]{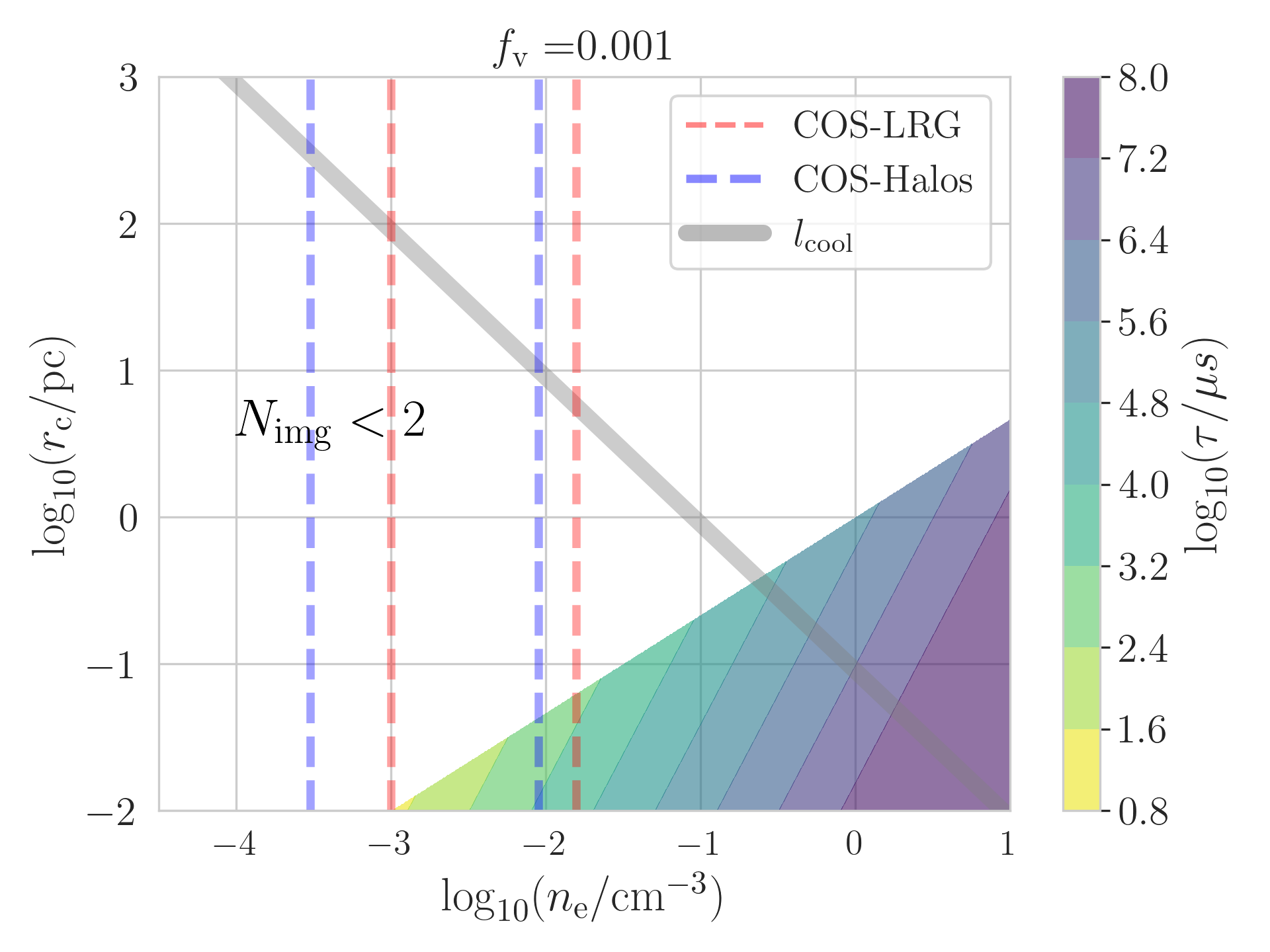}
\caption{Refractive scattering time delay produced by a halo composed of an ensemble of 
single-size clouds of radius \rc,
electron density $n_{\rm e}$, and two volume filling fraction values following Equations~\ref{eq:prop} and \ref{eq:propn}. We fix the 
parameters $\zell=0$, $r_{200}=200\, {\rm kpc}$, 
$d_{\rm s}/\dls=2$, $\dell \dls
/d_{\rm s} = 1 \, {\rm Gpc}$, and an observed frequency of 1 GHz. The white regions indicate 
the parameter space where refraction in the halo produces less than two images and, thus, does not result in scattering.
The vertical blue and red dashed lines express the 
range of electron density values 
inferred by \cite{Werk2014} 
and \cite{Prochaska2017} from COS-Halos 
data at $z \sim 0.2$, and  from 
the COS-LRG sample at $z \sim 0.4$ in \cite{Zahedy2019}, respectively. The  thick gray diagonal lines 
represent the shattering scale from 
\cite{Mccourt2018}. Overall, refraction 
from the bulk of low-density cool CGM gas does not  contribute 
to scattering. Dense gas clumps, however, produce strong 
scattering that may result in the suppression of  Milky Way scintillation.}\label{fig:constraints}
\end{figure}

Figure \ref{fig:constraints} shows scattering time delays induced by 
a halo composed of 
clouds of size and electron density fluctuations 
over ranges that cover frequently assumed halo values at low redshift
\citep[e.g.,][]{Faerman2023}, and with the formalism 
just described above. The panels show two values for the volume 
filling fraction, for comparison, noting that the highest value 
(top panel) is favored  by the observations of \citealt{Werk2014} 
(i.e., $f_{\rm v}\sim 10^{-1} - 10^{-2}$). 
Because we are most interested in the dependence on these parameters 
here, we have set the values $\zell=0$\footnote{In reality, the lens will 
always be at $z>0$, but we have assumed this value here for simplicity.}, $r_{200}=200\, {\rm kpc}$ (typical of a Milky Way halo mass 
$\log (M_{\rm h}/{\rm M_\odot}) \approx 12.2$ at $z=0$, although the 
cool gas might be contained within a slightly shorter radius), $\nu=1\,{\rm GHz}$, and the terms 
$d_{\rm s}/d_{\rm ls}=2$ (i.e., the lens is one-half the distance to the source, which 
maximizes the geometrical term) and $\dell \dls/d_{\rm s} = 1 \, {\rm Gpc}$. 
The white regions illustrate the parameter space where refraction produces less than two 
images and, therefore, there is no scattering, scatter broadening or an effect on scintillation. 
The blue and red dashed vertical lines  delimit the ranges of density  values 
inferred for $L^\star$ galaxies at $z\sim0.2$ from COS-Halos by \citealt{Werk2014} and 
\citealt{Prochaska2017}, and for massive quiescent galaxies at $z\sim 0.4$ 
from  COS-LRG by \citealt{Zahedy2019} (the red lines limit the region containing 68\% 
of the  volume density values estimated from their sample; see also 
\citealt{Chen2023})\footnote{The cloud sizes and densities may be linked to the 
volume filling fraction, but given the large uncertainties in 
these parameters we here treat them as independent from each other for visualization.}, 
respectively.
Finally, 
the thick gray diagonal lines represent the minimum cooling (shattering) cloud size  
proposed by \cite{Mccourt2018}, assuming a fully ionized gas. This scale 
depends on density as \citep[][their equation 15]{Faerman2023}
\begin{equation}\label{eq:ss}
    l_{\rm cool} \approx 9 \,{\rm pc} \left(\frac{n}{10^{-2}\,{\rm cm^{-3}}}\right)^{-1} ~,
\end{equation}
for a typical cool gas temperature of $10^4\,{\rm K}$, and where $n$ is the total gas 
density.

Overall, Figure~\ref{fig:constraints} illustrates two important results: First, 
the CGM gas covering most of the assumed parameter space 
(large, low density clouds)
produces no scattering because its refraction does not produce two 
images. 
Second, the halo may produce substantial scattering and
suppress scintillation only if the cloud size at CGM densities of $n_{\rm e} \lsim 
10^{-2}\,{\rm cm^{-3}}$ is $r_{\rm c} \lsim 0.2\,{\rm pc}$. 
Such gas clouds, however, would be 
approximately 100 times smaller than the shattering scale. 
If one adopts the shattering scale as a lower limit
to the cloud size at a given density, then 
scattering by the halo (where $N_{\rm img}\ge2$) would require pockets 
of gas at high densities, $n_{\rm e} \gtrsim 10^{-1}\,{\rm cm^{-3}}$, 
which may indeed coexist with the low-density gas in the CGM  
\citep[e.g.,][]{Chen2010,Rubin2011,Putman2011,Nielsen2013,Lan2017}.
At these high densities, the halo produces multiple ($10 - 100$; Equation~\ref{eq:imagetau}) images and 
the time delays are $\tau_{\rm s} \gtrsim 10 \,{\rm ms}$, in agreement with the 
results by \citealt{Jow2023}, but 
the detection of such highly scattered FRBs may be challenging. 
Furthermore, at the highest  
scattering times, the time delay between images may be larger than the pulse duration and  thus these may appear as separated pulses, adding to the  
complexity for detecting and characterizating such FRBs.

In summary, refraction by cool 
gas clouds with densities $n_e < 10^{-2} \, \rm cm^{-3}$
in halos will produce only a single image (\nimg=1)
and therefore not result in scattering.  
 No additional constraint
may be extracted for such gas from FRB scintillation.
However, a population of high density clumps embedded in the halo may yield $\nimg > 1$ and result in the significant scattering. 
We  assess the scenario of 
these dense clumps in more detail in Section~\ref{sec:dense}.

Figure~\ref{fig:100mhz} in the Appendix shows the same results as 
Figure~\ref{fig:constraints} but at 400 MHz, the minimum frequency of the CHIME 
experiment \citep[][]{Amiri2018}, for comparison and to visualize the effect 
of the strong dependence on frequency by, both, the number of images and the 
time delay. Overall, 
there is a slight increase in the parameter space where $N_{\rm img}\ge2$, as well as  higher time delays. Owing to the strong $\nu^{-4}$ frequency dependence of the number of images, 
Equation~\ref{eq:imagetau}  yields about 2 (100) images at a frequency of 
400 (100) MHz, with time delays of $\tau_{\rm s}\approx 0.1 (24)\,{\rm s}$ 
(Equation~\ref{eq:tauimage}), which possibly makes these highly-scattered 
FRBs challenging to detect at these frequencies.

\section{Physically-motivated Halo Gas Models}\label{sec:ensemblediff}

We explore now physically-motivated halo models and compare them 
to our previous results. In particular, we  make use of estimates of the dispersion measure in 
local halos in Section~\ref{sec:empirical}, and consider the case of a distribution of cloud sizes in 
Section~\ref{sec:analytical}.  Dense gas clouds are addressed 
in Section~\ref{sec:dense}, and the constraining power from the presence of 
scattering in high-redshift proto-cluster structures is 
investigated in Section~\ref{sec:highz}.  We finally apply our 
refractive formalism to two science cases in Sections~\ref{sec:faber24} and \ref{sec:p19}. 

\subsection{An Empirical Halo Model}\label{sec:empirical}

\cite{XYZ2019} examined a variety of empirical and numerical models 
that parameterize the ionized gas in local galaxy halos and, assuming a Milky 
Way halo mass and a single-component model for the halo gas, they showed that the electron 
density radial profiles from most models broadly agree to within one 
order of magnitude  (their figure 1). In particular, the models show values 
for the electron number density within the range $n_e= 10^{-3} - 10^{-4} \, {\rm cm^{-3}}$, 
from a few tens to about $150\,{\rm kpc}$ from the center of the halos, and 
electron columns on the order of ${\rm DM}=100\,{\rm pc\, cm^{-3}}$. We incorporate in 
our formalism the dispersion measure observable, commonly used in the FRB field, through 
\begin{equation}
    {\rm DM} = N_{\rm c}\, {\rm DM}_{\rm c} = 3 \fv \rtwo n_{\rm e}~,
\end{equation}
where ${\rm DM}_{\rm c} = 2\, r_{\rm c}\,  n_{\rm e}$ is the dispersion measure of one cloud intersected through its center. 
We can now  rewrite equations \ref{eq:tauimage}  and \ref{eq:imagetau} with this parameter and 
solve for the maximum cloud size that produces two images, resulting in 
\begin{align}\label{eq:rminimg}
    r_{\rm c}^3  \approx  \frac{(0.12\, {\rm pc})^3}{(1 + \zell)^{4}} &
    \left(\frac{\dell \dls/d_{\rm s}}{1\,{\rm Gpc}}\right)^2 \nonumber \\
    &\times \left(\frac{{\rm DM}}{100\,{\rm pc\,cm^{-3}}}\right)
    \left(\frac{ n_{\rm e}}{10^{-2} \, {\rm cm^{-3}}}\right)
    \left(\frac{1}{N_{\rm img} - 1}\right)
    \left(\frac{\nu}{1 \, {\rm GHz}}\right)^{-4} ~, 
\end{align}
with a corresponding refractive scattering time delay of 
\begin{align}\label{eq:taudm}
    \tau_{\rm s}  \approx  \frac{7}{(1 + \zell)^{3}} {\rm ms} &
    \left(\frac{\dell \dls/d_{\rm s}}{1\,{\rm Gpc}}\right)
    \left(\frac{{\rm DM}}{100\,{\rm pc\,cm^{-3}}}\right) \nonumber \\
    & \times   \left(\frac{ n_{\rm e}}{10^{-2} \, {\rm cm^{-3}}}\right) 
    \left(\frac{r_{\rm c}}{0.12 \, {\rm pc}}\right)^{-1} 
    \left(\frac{\nu}{1 \, {\rm GHz}}\right)^{-4}  ~ \hspace{0.8cm} \textrm{ if \hspace{0.8cm} $N_{\rm img}\geq2$} ~, \nonumber \\
\end{align}
Here above we have considered $ n_{\rm e} = 10^{-2} \, {\rm cm^{-3}}$, which is 
higher than the values estimated by \cite{XYZ2019}. This is because these authors considered a  
single (hot) component that dominates the gaseous halo mass, while the cool, ionized gas that 
we are interested in is expected to have higher densities
(e.g., in pressure equilibrium).
We have adopted the DM value 
as is in \cite{XYZ2019}, however, because their value is consistent with those inferred for 
intervening halos by \cite{Ilya2024}. 
Our results scale with the values of these parameters as expressed by the above equations.

Consistent with the results in the previous general case, with this set of parameters 
we overall find that the cloud sizes must be smaller than $r_{\rm c} \lsim 0.1\,{\rm pc}$  for 
the halo to produce multiple refractive images and scattering at millisecond scales.

\subsection{A Distribution of Cloud Sizes}\label{sec:analytical}

Instead of an ensemble of single-size clouds, we here adopt a distribution of cloud sizes 
following the results by \cite{Gronke2022} and \cite{Tan2024}. These authors 
performed simulations of survival and growth of gas clouds in multiphase media  
resembling the CGM, and found that the mass distribution of clouds can be described 
by a power law of the form ${\rm d}n/{\rm d}m \propto m^{-2}$, where $n$ now denotes 
the number of clouds in the distribution, down to the shattering 
scales of \cite{Mccourt2018}. By assuming a spherical shape and the same 
density for all of the clouds, we can rewrite the previous distribution with the cloud 
radius as $\frac{{\rm d}n}{{\rm d}r_{\rm c}} = \frac{{\rm d}n}{{\rm d}m}\frac{{\rm d}m}{{\rm d}r_{\rm c}}  \propto r_{\rm c}^{-4}$, which we adopt for our model. 

For this continuous distribution, the scattering angle now becomes (see the full derivation 
in Section~\ref{sec:deriv}),
\begin{equation}
    \alpha^2 \approx g_{\rm s}^2\int_{N_{\rm c}} \alpha_{\rm c}^2 = g_{\rm s}^2\frac{9}{4}  \fv \frac{r_{200}}{r_{\rm c, min}} \Delta m^2 ~ ,
\end{equation}
where $r_{\rm c, min}$ is the minimum cloud size in the distribution. Here we 
have ignored the terms containing the maximum cloud size, owing to the steep slope of 
the distribution and assuming that the two limits differ by a substantial  
($\gtrsim 1$ dex) amount. 
With this expression, the scattering time delay is now
\begin{equation}
    \tau_s \approx \frac{1}{(1+z_{\rm l})^3}\frac{d_{\rm l}d_{\rm ls}}{ d_{\rm s}} g_{\rm s}^2 \frac{9}{8c} f_{\rm v}\frac{r_{200}}{r_{\rm c, min}} \Delta m^2  ~.
\end{equation}
This equation is equivalent to Equation~\ref{eq:prop} above,  but replacing 
the cloud size there for the minimum cloud size in the 
distribution here.  The preceding factors, however, are not equal in these expressions   
for the two halo models. Because some of the clouds are larger than the minimum cloud size 
in the distribution scenario, the total number of clouds is smaller than in the case 
where all the clouds have the same size and this size equals the minimum cloud radius. This, 
in turn, reduces the number of transitions between media for the radio waves and, thus, 
the scattering time. Overall, the distribution of cloud sizes results in a reduction of 
the scattering time delay by a factor $\left(\frac{\beta-1}{\beta}\right)$ compared to 
the single-size case (Section~\ref{sec:deriv}), 
where $\beta=4$ is the fiducial power-law index of the distribution adopted here. 
For the number of images in the scenario with a 
size distribution, 
we have that (Section~\ref{sec:deriv}), 
\begin{equation}
    N_{\rm img} -1 \approx \frac{1}{(1+\zell)^4} \left(\frac{\dell \dls}{d_{\rm s}}\right)^2 \Delta m^2 
    \,\frac{3}{2}\, \fv \, \frac{\rtwo}{r_{\rm c, min}^3} ~,
\end{equation}
when $\beta = 4$, corresponding to a reduction in the number of images by a factor 
$\left(\frac{\beta -1}{\beta +2}\right)$ compared to the single-size case. For smaller 
values of $\beta$, 
the differences between the two halo models grow larger, as expected from having a flatter distribution 
of sizes, i.e., an increased fraction of large clouds. We recall 
that these expressions are  
only valid for steep distributions; one should take into account the terms containing 
the maximum cloud size for the case of distributions with small $\beta$ values.

In summary, for a halo consisting of a (steep) distribution of cloud 
sizes following a power law with index $\beta$, the time delay and number of images are reduced 
by factors $\left(\frac{\beta-1}{\beta}\right)$ and $\left(\frac{\beta -1}{\beta +2}\right)$, respectively, 
compared to the single-size cloud case where this cloud size equals the minimum size of the clouds in the 
distribution. For $\beta=4$ obtained from simulations, these terms correspond to factors $3/4$ and $1/2$, 
respectively, the latter further contributing to the small number of images expected for the bulk  
of CGM gas.

\subsection{Dense Gas Clumps Near Galaxies}\label{sec:dense}

In previous sections we found that when adopting the low density values 
characteristic of cool gas in the circumgalactic medium of  
low-redshift halos, refraction from one halo generally does not 
create two images and, therefore, does not produce scattering. A fraction of dense $n_{\rm e}\gtrsim 0.1\,{\rm cm^{-3}}$ 
gas, however, may be expected to also co-exist in the central regions of the CGM and, as illustrated in 
Figure \ref{fig:constraints},  intersecting a number of these dense clumps\footnote{We use the word clump instead of cloud here to 
differentiate these high 
density media from the ones considered before.} can   
result in substantial scattering. 

We now estimate the probability of finding this dense ($n_{\rm e}\gtrsim 0.1\,{\rm cm^{-3}}$) gas  along FRB sightlines. Specifically, the probability 
of intersecting dense clumps can be written as 
\begin{equation}\label{eq:nfc}
    p_{\rm cl} = {\bar{n}_{\rm d}}f_{\rm c} ~,  
\end{equation}
where $f_{\rm c}$ is the areal covering fraction of dense (clumpy) gas in  
a region within a radius from the center of a halo, and 
\begin{equation}\label{eq:num}
   {\bar{n}_{\rm d}} (z_{\rm FRB})= \int_0^{z_{\rm FRB}} (1 + z)^2\,d_{\rm H}(z) \int \pi r_d^2\, \frac{{\rm d}n_{\rm h}}{{\rm d}M_{\rm h}}(z)\, {\rm d}M_{\rm h}\,{\rm d}z ~
\end{equation}
is the mean number of regions hosting dense clumps along the sightline of 
an FRB at redshift $z_{\rm FRB}$ \citep{Padmanabhan2002}. Here, 
${\rm d}n_{\rm h}/{\rm d}M_{\rm h}$ is the comoving number density of halos 
of mass $M_{\rm h}$, and the term $d_{\rm H}(z)= {c}\, {\rm H_0}^{-1} [\Omega_\Lambda + \Omega_{\rm m}(1+z)^3]^{-1/2}$, with 
${\rm H_0}$, $\Omega_{\Lambda}$, and $\Omega_{\rm m}$ denoting the Hubble constant, and the 
dark energy and matter densities today, respectively. The parameter 
$r_d = 10$ kpc determines the radius out to which the dense  gas may 
be expected to reside.  This distance is motivated by the fact that we do not 
expect clouds with $n_{\rm e}\gtrsim 0.1$ cm$^{-3}$ -- the approximate density 
at which the shattering scale becomes small enough for multiple images -- at 
significant distances from Milky Way like galaxies given that photoionized gas 
in the interstellar medium has  $n_{\rm e}\sim 1$ cm$^{-3}$.  Additionally, as 
noted in Section~\ref{sec:intro}, $r_d = 10$ kpc corresponds to typical distances to complexes of high velocity clouds \citep{Wakker2008}, estimated to often 
have densities of  $n_{\rm e}\sim 0.1$ cm$^{-3}$ like those considered here.

\begin{figure}\center 
\includegraphics[width=0.8\textwidth]{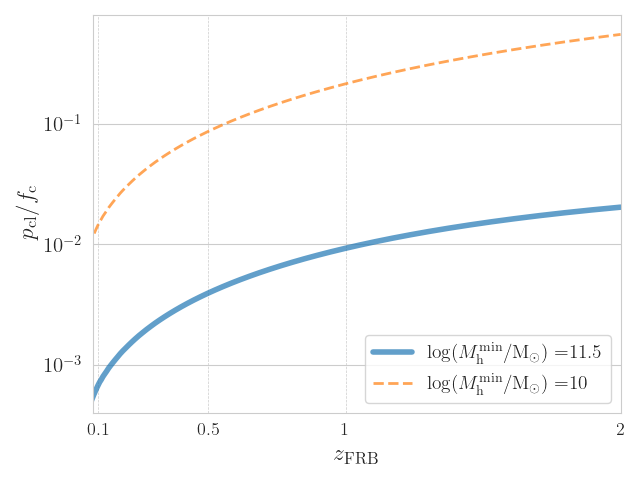}
\caption{Probability of intersecting dense  ($n_{\rm e}\gtrsim 0.1\,{\rm cm^{-3}}$) clumps within the inner 10 kpc of intervening halos along 
the sightline of a redshift $z_{\rm FRB}$ FRB, which may result in refractive scattering, over the (unknown) covering fraction of this gas in such a central region.   
The solid blue line denotes  halos of mass $\log(M_{\rm h}/{\rm M_\odot}) \geq 11.5$, which are the most likely to host cold gas, and the dashed orange line 
represents halo masses $\log(M_{\rm h}/{\rm M_\odot}) \geq 10$ for comparison. Overall, a small fraction of low-redshift ($z\lesssim 0.5$) FRBs are expected 
to show refractive scattering from 
this dense gas given the small probabilities of intersecting such regions.}
\label{fig:hit}
\end{figure}

Figure~\ref{fig:hit} displays the probability of intersecting dense clumps 
along the sightline of an FRB at redshift $z_{\rm FRB}$ over the (unknown) 
covering fraction of such dense gas within the inner 10 kpc of halos\footnote{Although we do not write it explicitly for simplicity, we 
note that the covering fraction is expected to depend on halo properties and 
redshift, where over order-unity differences with redshift are likely to occur 
as we expect denser CGMs at higher redshifts 
\citep[e.g.,][]{Lan2017,Dutta2020,Lan2020}.}. 
This quantity simply corresponds to the mean number of  regions hosting dense gas along an FRB sightline (Equation~\ref{eq:nfc}). For these calculations, we 
have used the \citealt{Tinker2008} halo mass function from the 
\texttt{hmf} package \citep{hmf2013}, integrated up to the default upper 
halo mass limit of  
$\log(M_{\rm h}^{\rm max}/{\rm M_\odot}) = 15$, although this exact 
value has little impact on the results. The solid blue line denotes a minimum halo mass limit of  $\log(M_{\rm h}^{\rm min}/{\rm M_\odot}) = 11.5$, 
as halo masses near that of the Milky Way are the most efficient at forming stars and hosting cold gas. The dashed orange line corresponds to the case 
$\log(M_{\rm h}^{\rm min}/{\rm M_\odot}) = 10$, down to small dwarf galaxies, for comparison, although such small halos are unlikely to have cold dense gas.   Overall, we anticipate a small fraction of bursts showing refractive scattering from this 
dense gas. When this happens, it is likely that the FRB sightline passes close to a galaxy, and the images are expected to be separated by milliseconds, if not longer, 
at $\nu\sim 1$ GHz  (Equation~\ref{eq:tauimage}).

\subsection{$z\gtrsim 2$ Proto-clusters around Quasars}\label{sec:highz}

During the last few decades, there have been numerous
detections and detailed studies of 
large (hundreds of kpc across) and dense ($n_{\rm H}\sim 1\,{\rm cm^{-3}}$) 
environments at redshift $z\gtrsim2$ with overdensities suggesting they constitute 
the cool gas component of the progenitors 
of present-day galaxy clusters 
\citep[e.g.,][]{Steidel2000,Cantalupo2014,Hennawi2015,Geach2016,Cai2018,Li2024}. 
We explore here the gas constraints that can be obtained from the scenario of an FRB behind one of 
these proto-clusters. 

The discovery of one of these massive structures at $z\approx2$ was reported by 
\cite{Hennawi2015}. These authors examined the nebular emission from the gas, as well as the absorption 
signature imprinted by this gaseous environment in the spectrum of a background quasar at an impact parameter of 
$176\,{\rm kpc}$ from the center of the proto-cluster. From photoionization 
modeling, \cite{Hennawi2015}
inferred hydrogen volume and column densities of $n_{\rm H}\approx 2\,{\rm cm^{-3}}$ 
and $\log (N_{\rm H}/{\rm cm^{-2}}) = 20.4\pm0.4$, respectively, and an ionization fraction of 
$\sim 90\%$. 

Given the large ionization fraction inferred by \citep{Hennawi2015}, for our 
calculations we simply assume  $\log (N_{\rm e}/{\rm cm^{-2}}) = 20.4 \pm 0.4$, 
which corresponds to ${\rm DM \approx 84^{+126}_{-51}\, pc\,cm^{-3}}$, and $n_{\rm e}= 2\,{\rm cm^{-3}}$, 
as well as that all this gas (DM) contributes to scattering. 
Introducing these values into Equation~\ref{eq:rminimg}, the limiting (maximum) 
cloud size value to obtain at least two images is $r_{\rm c}\approx 0.15^{+0.05}_{-0.04}\,{\rm pc}$, which 
yields a time delay at $1$~GHz of $\tau_{\rm s}\approx 34^{+15}_{-10}\,{\rm ms}$. These two values are of the same order 
as those we obtained for the low-redshift cases despite the high gas density involved in this scenario, 
due to the strong redshift dependencies of the number of images and time delay. However, contrary to the low-redshift case, 
scattering from high-$z$ proto-cluster cases may be used to probe the shattering scales, because 
the maximum cloud size required for scattering found here is similar to the shattering scale expected at these high densities (Equation~\ref{eq:ss}).

Similar to the previous section, we can further assess the probability of 
intersecting  proto-cluster structures along an FRB sightline. By comparing 
the sizes of the observed area in \cite{Hennawi2015} and around 29 quasars in \cite{Hennawi2013}, as well as  
the extent to where the emission from the proto-cluster gas was detected, \cite{Hennawi2015}  
argued that about   $f_{\rm qso}=10\%$ of  $z\sim2$ quasars may have extended 
gas structures like the one discussed here, with  a gas covering fraction of $f_{\rm c,proto}\sim 30\%$ out to  $r_{\rm proto}\sim 150$ kpc \citep[section 7 in][]{Hennawi2015}. 
Furthermore, by adopting the quasar luminosity function of \cite{Hopkins2007} down to an apparent magnitude limit of $V=24$, as well as 
an obscuration fraction value of 50\% consistent with \cite{Reyes2008} and \cite{Lusso2013}, \cite{Hennawi2015} derived a mean comoving number density of quasars at $z\sim 2$ of $n_{\rm qso} = 3.8\times 10^{-5}\, {\rm cMpc}^{-3}$. 
With this information we can now write the probability of intersecting (at least) 
one proto-cluster along an FRB sightline as \citep{Masribas2016}  
\begin{equation}\label{eq:probproto}
    p_{\rm proto} = 1 - (1 - f_{\rm c,proto})^{\bar{n}_{\rm proto}} ~, 
\end{equation}
where $(1 - f_{\rm c,proto})^{\bar{n}_{\rm proto}}$ is the probability of not 
finding any proto-cluster in any of the ${\bar{n}_{\rm proto}}$ quasars 
hosting them, and where the probability of intersecting the proto-cluster in 
one host quasar is given by the areal covering fraction of the proto-cluster 
gas, $f_{\rm c, proto}$. Following Equation~\ref{eq:num}, the mean number 
of quasars hosting proto-clusters is 
\begin{equation}\label{eq:nproto}
   {\bar{n}_{\rm proto}} (z_{\rm FRB})=  \pi r_{\rm proto}^2 f_{\rm qso} n_{\rm qso} \int_2^{z_{\rm FRB}} (1 + z)^2 d_{\rm H}(z){\rm d}z ~,
\end{equation}
where we assume that all proto-cluster structures are well-characterized  
by the  aforementioned $z\sim 2$ parameter values in any case, including the mean number density of quasars, and that proto-clusters exist only at $z\geq 2$. 
Solving the above equations we find a probability of intersecting a proto-cluster 
with an FRB sightline of $p_{\rm proto}\approx 10^{-3}$ 
($p_{\rm proto}\approx 10^{-4}$) at $z_{\rm FRB}\approx3$ ($z=2.1$). This indicates  
that a future large ($\gtrsim 1000$) sample of $z\gtrsim 3$ FRBs may enable 
probing the shattering scale and setting further constrains on 
high-redshift CGM gas.

\subsection{Science Case 1: FRB20221219A}\label{sec:faber24}

\cite{Faber2024} recently reported the discovery of the potentially highly scattered FRB20221219A at 
$z= 0.55$, whose ``overcrowded'' sightline traverses two extragalactic halos and the 
outskirts of a galaxy cluster. These authors 
measured a scattering timescale from the pulse broadening 
of $\tau = 19.2\,{\rm m s}$  
at 1.4 GHz, although they were not able to determine if it had the characteristic $\sim \nu^{-4}$ frequency dependence to help distinguish scattering from the intrinsic width of the FRB. They argued, based on the observed large scattering timescale, that an intervening CGMs was more 
plausible to be the source of 
scattering than 
the Milky Way or the FRB host, but scintillation measurements that could 
confirm that CGM picture were not obtained. 

As an application of the formalism developed above, we can now examine the potential contribution of 
refraction from the intervening halos to scattering.  At first glance, such a large scattering time \emph{could} be produced by refractive 
scattering\footnote{We note that considering our fiducial 1 GHz frequency 
yields $\tau_{\rm s}\approx 74$ ms, and that, specifically, 
this scattering timescale corresponds to the $1/e$ value from the pulse 
broadening, while its tail reaches larger times.}. However, the geometric term and the DM halo values 
in \cite{Faber2024} for FRB20221219A are  both a factor of about ten smaller than those 
considered in Equations \ref{eq:rminimg} and \ref{eq:taudm}.  For refraction to be 
able to produce a scattering time of the order of that measured,
Equation~\ref{eq:taudm}  results in the condition $n_{\rm e}/r_{\rm c}\sim 10 \,{\rm cm^{-3}\, pc^{-1}}$ which, applied to Equation~\ref{eq:rminimg} yields a 
maximum cloud size of $r_{\rm c}\sim 0.1 \,{\rm pc}$ to produce two 
images. Although these results are consistent with shattering scale clouds of 
$n_{\rm e}\approx 1 \,{\rm cm^{-3}}$, gas of such densities is unlikely to 
inhabit the measured distances of 36 kpc and 43 kpc from the center 
of the intersected halos as we discussed previously. Smaller-size 
clouds could produce more images and would require lower gas densities but 
this would imply cloud scales far below the shattering size. Overall, 
refractive scattering appears as an unlikely driver of this signal.

\subsection{Science Case 2: FRB20181112}\label{sec:p19}

Contrary to the large scattering time discussed above, we consider now 
FRB20281112, detected by the Commensal Real-time Australian Square 
Kilometer Array 
Pathfinder (ASKAP) Fast Transient survey 
\citep[CRAFT;][]{Bannister2019,Cho2020,Scott2023} and analyzed in 
\cite{Prochaska2019}. This FRB shows a total of DM$=589.27\,{\rm pc\,cm^{-3}}$, a scattering time of $\tau\approx 40\,\mu{\rm s}$, and 
a host at $z=0.4755$. Its sightline intersects a galaxy at $z=0.3674$, 
at an impact parameter of 29 kpc, giving a DM$=50-120\,{\rm pc\,cm^{-3}}$. 
Adopting DM$=100\,{\rm pc\,cm^{-3}}$ and 
plugging these values into Equation \ref{eq:taudm} yields the relation 
$n_{\rm e}/r_{\rm c}\sim 4\times 10^{-3} \,{\rm cm^{-3}\, pc^{-1}}$ that, combined with 
Equation~\ref{eq:rminimg} results in a maximum cloud size of 
$r_{\rm c}\sim 4\times 10^{-3} \,{\rm pc}$. These values arise from 
the fact that the measured scattering time is very small compared to the 
millisecond values obtained in our fiducial calculations; the cloud size must 
be small to produce two images, but to obtain small scattering times the 
density needs 
to be even smaller to counterbalance the effect of the cloud size.   
This places the gas in an unlikely region of the CGM parameter space (beyond 
the bottom left corner of Figure~\ref{fig:constraints}), 
suggesting that refraction \emph{is not} the main driver of this scattering. 

The two cases just discussed above serve as a proof of concept and example 
applications of our formalism. We are working on statistically applying this 
approach to a large number of FRBs to connect the fraction of FRBs without 
scintillation to the average properties of dense gas in halos.

\section{Refractive vs Turbulent CGM Scattering}\label{sec:turb}

Turbulence in the CGM may act as another source of scattering that confuses the determination of whether refractive scattering is happening.  Furthermore, if turbulence is occurring within the parsec-scale clouds that can produce refractive scattering, it would increase the scattering in small clouds relative to our refractive calculation.  Here we consider how the strength of turbulent scattering compares to our refractive scattering in the limit where the turbulent cascade scales in the manner of Kolmogorov turbulence to sub-astronomical unit lengths which, as we discussed in Section~\ref{sec:intro}, requires specific circumstances in order to hold \citep[e.g.][]{Lithwick_2001, Prochaska2019}.

For subsonic driving with Mach number ${\cal M} < 1$, as is most likely in the CGM (and CGM cloud velocity widths limit ${\cal M} \lesssim 2$; \citealt{2024arXiv240603553F}), the variance of density fluctuations from isothermal turbulence scales as ${\cal M}^2$ on the driving scale, with an ${\cal O}(1)$ coefficient that depends on the properties of the driving \citep{1994ApJ...423..681V}\footnote{Adiabatic turbulence scales as ${\cal M}^3$ and hence the fluctuations are slightly smaller.  Adiabatic is more likely (1) for the virialized rather than cold CGM gas or (2) if the driving scale is smaller than the shattering scale.}.  The scattering time can be related to the length scale where the phase structure function equals unity, called the diffractive scale, as $\tau_s = \lambda/(2\pi c) (r_F/r_{\rm diff})^2$, where $r_F$ is the Fresnel scale  \citep[c.f. section 2.3 in][for definitions and where redshift factors appear]{Prochaska2019}.  The phase structure functions from Kolmogorov turbulence at a fixed column density 
$N_{\rm e}$ in cold gas up to physical constants and an ${\cal O}(1)$ coefficient equals $N_{e, d}^2 {\cal M}^2 (r/L)^{5/3}$ for diffraction, whereas the structure function of refractive fluctuations from cloudlets, $N_{e, r}^2 (r/r_c)^{2}$, where the quadratic scaling applies for most cloudlet geometries (c.f. \citealt{Vedantham2019} and \citealt{Prochaska2019}). As the total column participating in turbulence $N_{e, d}$ may be different than in the refractive cloudlets $N_{e, r}$, we can define their ratio $R \equiv N_{e, r}/N_{e, d}$.  Thus, for turbulence and refractive lensing to produce the same scattering time, the ratios of the structure functions at $r_{\rm diff}$ must be equal:
\begin{equation}
\frac{R^2}{{\cal M}^2} \left( \frac{L}{r_c}\right)^{5/3} \left( \frac{r_{\rm diff}}{r_c} \right)^{1/3} = 1 ~. 
\label{eqn:structurefunctionratio}
\end{equation}
 Let us assume a similar column participating in turbulence as refractive scattering, so $R=1$, so that our relation can be rewritten as $L/r_c/{\cal M}^{6/5} = ({r_c}/r_{\rm diff})^{1/5}$.   For millisecond scattering times at cosmological distances, $r_{\rm diff} \sim 10^{12}$~cm \citep{Prochaska2019, Ocker2025}.  We adopt this value, although our conclusions depend weakly on this choice. If  to evaluate the $r_{\rm diff}/r_c$ factor we further take $r_c = 0.03~$pc, close to what is required for refraction at CGM densities -- although we note that the dependence of our conclusions on $(r_{\rm diff}/{r_c})$ is quite weak --, then equation~\ref{eqn:structurefunctionratio} reduces to $L/r_c/{\cal M}^{6/5} = 10$.  This means that the scattering time delay from turbulence is the same as from refractive scattering by cloudlets when the driving scale of the turbulence is $L\approx 10\, r_c$ if ${\cal M} = 1$ or when $L\approx r_c$ if ${\cal M} = 0.1$

We can further put a constraint on how much scattering can be enhanced by turbulence by considering the energetics to drive such turbulence throughout all the cold CGM gas.  The dissipation time is the turnover time for the largest eddies, which we are setting to be the cloud size. Thus, the energy density dissipated is roughly equal to:

\begin{equation}
\epsilon = \left(\frac{1}{2}\rho v^2\right) \frac{v}{L},
\end{equation}
where $v$ is the velocity of eddies on the driving scale $L$. If the CGM is composed of small clouds, the natural driving scale is the cloud scale so $L \sim r_c$.

If the turbulence is driven with Mach number ${\cal M}$ at the cloud scale, this becomes:

\begin{equation}
\epsilon = \left(\frac{1}{2}\rho  {\cal M}^2 c_s^2  \right) \frac{{\cal M} c_s}{r_c}~,
\end{equation}
where $c_s$ is the sound speed.

The total energy dissipated in the CGM is then
$E_{\mathrm{turb}} = \frac{1}{2} {\cal M}^3 \rho c_s^3 V_{c} r_c^{-1}$~, 
where $V_c$ is the total volume of the clouds,
and this needs to be less than the energy in galactic feedback. If galactic feedback is dominated by supernova, each supernova has kinetic energy $E_{\mathrm{SN}} = 10^{51}$ erg and it takes 100 solar masses of stars formed to have one supernova:

\begin{equation}
E_{\mathrm{feedback}} = \frac{E_{\mathrm{SN}}\, \mathrm{SFR}}{100}  = 10^{49} \, \mathrm{SFR} \,\, \mathrm{erg~yr^{-1}}~,
\end{equation}
where SFR is the star formation rate in solar masses per year.
A Milky Way-like galaxy forms about one solar mass a year. If we say the radius of cold gas in the CGM is 100 kpc, consistent with observations \citep{Werk2016, Faerman2024}, the volume filling fraction of cold droplets of size $r_{\rm c}$ is $f_{\rm v}$,  and the clouds have a density of $\rho = 1.2\, m_{\rm p}\, n_{\rm e}$, where $m_{\rm p}$ is the proton mass and $1.2$ the mean molecular weight, then the dissipated energy is:

\begin{equation}
E_{\mathrm{turb}} = 8 \times 10^{49} \,\, \mathrm{erg~yr^{-1}} \left(\frac{f_{\rm v}}{10^{-3}}\right) \left(\frac{n_{\rm e}}{10^{-2} \,\, \mathrm{cm}^{-3}}\right) \left(\frac{r_{\rm c}}{1 \,\, \mathrm{pc}}\right)^{-1}  {\cal M}^3 \left( \frac{T}{1.5\times10^4\,{\rm K}} \right)^{3/2}~.
\end{equation}
For parsec-scale clouds and ${\cal M}\sim 1$, this exceeds the energy in feedback even for a rather small volume filling factor of $f_{\rm v} \sim 10^{-3}$ and a density like found in the low-$z$ CGM of $n_{\rm e} \lesssim 10^{-2}\,{\rm cm^{-3}}$.  We could adjust ${\cal M}$ to be smaller, but by ${\cal M} = 0.1$ the strength of diffractive scattering from turbulence is comparable to refractive scattering from the cloudlet size (Equation~\ref{eqn:structurefunctionratio}).

Thus, this energetics argument suggests that turbulence cannot enhance the scattering of parsec-scale clouds by a large factor over our refractive estimates when there are multiple refractive images. While driving turbulence on a larger scale than $r_{\rm c}$ is possible to avoid this argument, Equation~\ref{eqn:structurefunctionratio} shows that the driving scale of turbulence can only be an order of magnitude larger than our cloud sizes to produce scattering times similar to those produced by refractive scattering and then for ${\cal M} = 1$.  Since it is the existence of the clouds themselves that offers the cloud scale as a viable driving scale, this suggests that if CGM scattering times like those estimated in this paper are found, it would likely owe to a mist of clouds being present in the CGM, even if some of the scattering is diffractive.

\section{Conclusion}\label{sec:conclusions}

In this work we have assessed the contribution of extragalactic halos 
to refractive FRB scattering, and the  constraints on the physical properties 
of the halo gas  obtainable from this observable. Our main results can be 
summarized as follows:

\begin{itemize}
    \item[1.] Refraction from the bulk of low-redshift circumgalactic gas in extragalactic halos at densities $n_{\rm e} 
    \lsim 10^{-2}\,{\rm cm^{-3}}$ is not likely to impart scattering delays, and hence suppress the Milky Way scintillation, 
    because the sizes of photoionized clouds are likely too large to produce multiple images. To be able to create multiple images in the low-redshift CGM, the gas 
    clouds should be more than a hundred times smaller than the shattering scale predicted by \cite{Mccourt2018}.

    \item[2.] Considering the shattering scale as a lower limit for the sizes of 
    low-redshift CGM clouds, the production of scattering requires gas at densities of $n_{\rm e} \gtrsim 0.1\,{\rm cm^{-3}}$. In this case, 
    the refractive time delays are $\gtrsim 1 - 10$ ms.

    \item[3.] While scattering times of $\gtrsim 1 - 10$ ms  from low-redshift 
    CGM may complicate the detection of these FRBs owing to the pulse smearing, 
    these long delays may be a unique signature of refractive scattering: We 
    have argued via constraints on galaxy energetics that diffractive (turbulent) scattering from the CGM is unlikely to be able to produce significantly larger scattering times, and so a population of FRBs with very long scattering times would likely be indicative of sub-parsec CGM clouds.  Motivated by high-velocity clouds of similar densities ($n_{\rm e} \gtrsim 0.1\,{\rm cm^{-3}}$) that 
    reside within $\sim 10~$kpc from galaxies, we estimate that a small fraction 
    of FRB sightlines are likely to intersect and be affected by such a dense gas.

    \item[4.] At redshifts $z\sim 2 - 3$ massive (hundreds of kpc across) 
    proto-clusters observed around quasars are the environment most likely to produce scattering if they are inhabited by shattering-scale ($0.1$ pc) 
    cloudlets. These media contain substantial $n_{\rm e} \sim 1\,{\rm cm^{-3}}$ cool gas, for which shattering-scale clouds 
    yield tens of millisecond scattering delays. A fraction of $\sim 10^{-3}$ 
    high-redshift FRBs sightlines are expected to intersect such environments.

    \item[5.]   Our conclusions about scattering times and the potential for detecting refractive scattering are unchanged when considering empirical dispersion measure estimates for low-redshift halos, as well as when 
    adopting a distribution of halo cloud sizes motivated by numerical simulations.

\end{itemize}

In conclusion, refractive scattering in intervening low-redshift halos is 
sensitive to shattering scale clouds of density $n_{\rm e} \gtrsim 0.1\,{\rm cm^{-3}}$, which are only likely to impact a small fraction of FRB sightlines 
and then only in the picture of ample shattering-scale cloudlets. Clouds at 
the highest densities ($n_{\rm e} \gtrsim 1\,{\rm cm^{-3}}$) will 
have an even smaller effect since they 
are expected to reside in the innermost regions of the CGM. In an 
upcoming work, we investigate the constraints on such a very dense 
gas obtainable from the 
fraction of FRBs with suppressed scintillation in a large sample. More detailed 
calculations considering multiple screens would benefit from numerical simulations. 
However, simulations with enough resolution to treat the radiative transfer 
effect of scattering \citep[e.g.,][]{Pradeep2025} are not yet implemented within 
galaxy-scale codes, as this would require the accurate modeling of a large range of 
physical scales becoming computationally prohibitive. A first approach, beyond 
simple analytical calculations with more than one screen, may consider a combination 
of numerical radiative transfer calculations and the analytical modeling of halo 
gas. Given that 
the smallest scales of gas in halos are still under debate, a semi-numerical method 
may allow for extending calculations to multiple intervening halos efficiently, 
while also capturing the main physical effects in the CGM.

\section*{Acknowledgements}

L.M.R thanks the Radio Astronomy group at Caltech 
for kind hospitality while part of this work was 
performed. We are grateful to Kate Rubin for many discussions and 
performing CGM cloud simulations for our work. We also thank our colleagues 
Liang Dai, Jakob Faber, Stella Ocker, Andrew Robertson, Dylan Jow, Gwen 
Rudie, Ilya Khrykin, Mandy Chen, Mawson Sammons, Phil Hopkins, Sachin Pradeep, 
Cameron Hummels, Fakhri Zahedy, Michael Rauch, Adam Lanman, Nicolas Tejos, 
Robert Main, Kenzie Nimmo, Calvin Leung, and 
Clancy James, among others, 
for useful discussions. 
Authors L.M.R and J.X.P, as members of the
Fast and Fortunate for FRB Follow-up (F4) team, acknowledge
support from NSF grants AST-1911140, AST-1910471, and
AST-2206490.  M.M. acknowledges support from NSF grant 2007012.

\bibliographystyle{aasjournal}
\bibliography{frbscat}\label{References}


\appendix

\section{Cloud constraints at 400 MHz}\label{sec:100mhz}

\begin{figure}[h]\center 
\includegraphics[width=0.85\textwidth]{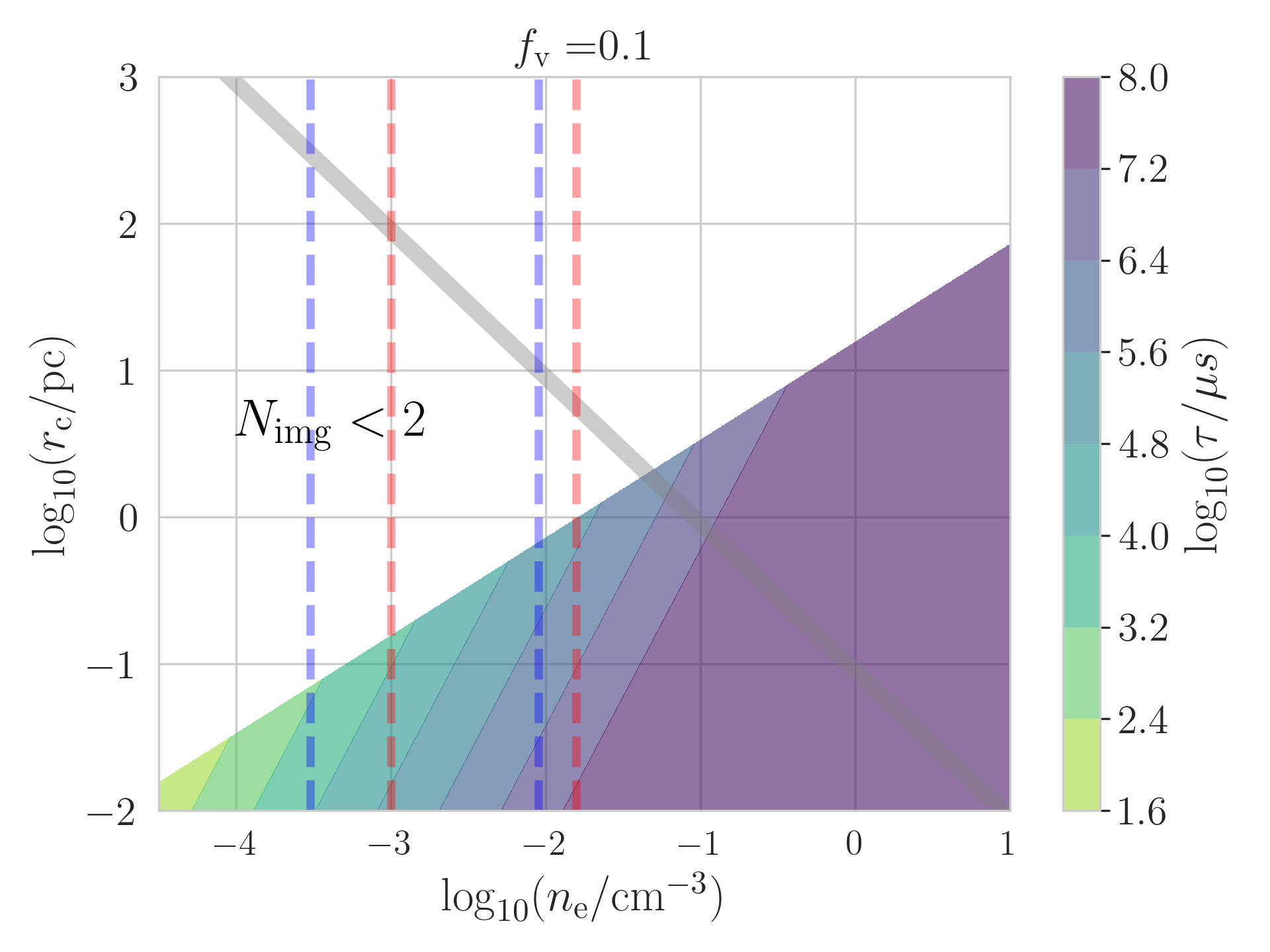}
\includegraphics[width=0.85\textwidth]{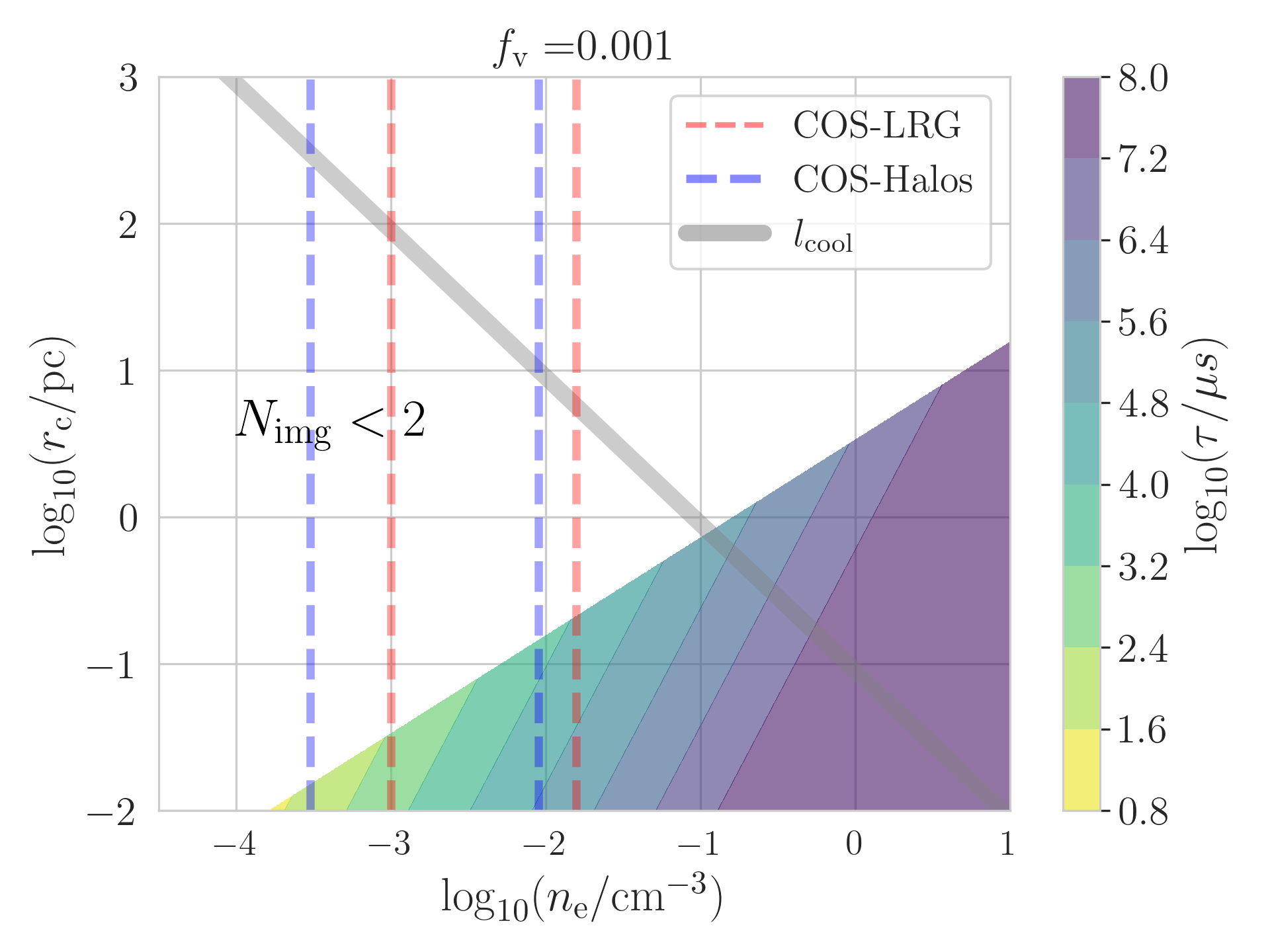}
\caption{Same as Figure~\ref{fig:constraints} but at a frequency of 400 MHz.}
\label{fig:100mhz}
\end{figure}

\section{Derivation of the scattering angle and number of images for a cloud-size distribution}\label{sec:deriv}

For a cloud-size distribution $\frac{{\rm d}{ n}}{{\rm d}r_{\rm c}}\propto r_{\rm c}^{-4}$, 
the scattering angle is 
\begin{align}
    \alpha^2 \approx  & \;g_{\rm s}^2{\int_{N_{\rm c}}\alpha^2_{{\rm c}} } = {2\,g_{\rm s}^2 N_{\rm c}}\, \Delta m^2 
    =  2 \,g_{\rm s}^2\Delta m^2 \frac{\int\frac{{\rm d}{ n}}{{\rm d}r_{\rm c}} 2\int_{0}^{\sqrt{r_{200}^2 -b^2}} \frac{\pi r_{\rm c}^2}{\frac{4}{3}\pi r_{\rm c}^3} f_{\rm v}\left(\sqrt{s^2+b^2}\right) \,{\rm d}{ s} \,{\rm d} r_{\rm c}}{\int \frac{{\rm d}{ n}}{{\rm d}r_{\rm c}}{\rm d} r_{\rm c}}    \nonumber \\
    \approx &\,  3\, g_{\rm s}^2 f_{\rm v} r_{200} \Delta m^2 \frac{\int \frac{{\rm d}{ n}}{{\rm d}r_{\rm c}} \frac{{\rm d} r_{\rm c}}{r_{\rm c}}}{\int \frac{{\rm d}{ n}}{{\rm d}r_{\rm c}}{\rm d} r_{\rm c}} = g_{\rm s}^2
    \frac{9}{4} f_{\rm v} \frac{r_{200}}{r_{\rm c, min}} \Delta m^2 ~, 
\end{align} 
where we have assumed that the term with the maximum cloud size is negligible 
compared to the term with the minimum cloud size.
For the case of a power law with arbitrary index $\beta$, we have
\begin{equation}
    \alpha^2 = g_{\rm s}^2\frac{3(\beta -1)}{\beta} f_{\rm v} \frac{r_{200}}{r_{\rm c, min}} \Delta m^2 ~.
\end{equation}

We can then compute the number of images for the distribution case as
\begin{align}
    N_{\rm img} -1 \approx N_{\rm c} k_{\rm c}^2  
    = & \frac{1}{(1+z_{\rm l})^4} \left(\frac{d_{\rm l} d_{\rm ls}}{d_{\rm s}}\right)^2 \Delta m^2 
    \,3\, f_{\rm v}\, r_{200}\frac{\int \frac{{\rm d}{ n}}{{\rm d}r_{\rm c}} \frac{{\rm d} r_{\rm c}}{r_{\rm c}^3}}{\int \frac{{\rm d}{ n}}{{\rm d}r_{\rm c}}{\rm d} r_{\rm c}}   \nonumber \\
    = & \frac{1}{(1+z_{\rm l})^4} \left(\frac{d_{\rm l} d_{\rm ls}}{d_{\rm s}}\right)^2 \Delta m^2 
    \,\frac{3}{2}\, f_{\rm v}\, \frac{r_{200}}{r_{\rm c, min}^3} ~, 
\end{align} 
and for an arbitrary index $\beta$ we obtain
\begin{equation}
    N_{\rm img} -1 \approx \frac{1}{(1+z_{\rm l})^4} \left(\frac{d_{\rm l} d_{\rm ls}}{d_{\rm s}}\right)^2 \Delta m^2 
    \,\frac{3(\beta -1)}{(\beta +2)}\, f_{\rm v}\, \frac{r_{200}}{r_{\rm c, min}^3} ~.
\end{equation}

\end{document}